\title{Integrable (3+1)-dimensional generalization for dispersionless Davey–Stewartson system} 
\author{Antonio J. Pan-Collantes \thanks{
                  Departamento de Matem\'aticas\\ 
                  Universidad de C\'{a}diz - UCA \\
                  Puerto Real\\
                  \texttt{antonio.pan@uca.es}}
        }
\documentclass{article}
\usepackage{graphicx}
\usepackage{tikz-cd}
\usepackage[utf8]{inputenc}  
\usepackage{hyperref}
\hypersetup{
	colorlinks=true,
	linkcolor=blue,
	filecolor=magenta,      
	urlcolor=cyan,
	citecolor=magenta,
}
\usepackage{graphicx}       
\graphicspath{{media/}}     
\usepackage{float}

\usepackage{amssymb, amsmath}
\usepackage{amsthm}
\theoremstyle{remark}
\newtheorem{remark}{Remark}

\theoremstyle{definition}
\newtheorem{theorem}{Theorem}

\newtheorem{proposition}{Proposition}

\begin{document}

\newpage

\maketitle

\begin{abstract}
This paper introduces a (3+1)-dimensional dispersionless integrable system, utilizing a Lax pair involving contact vector fields, in alignment with methodologies presented by A. Sergyeyev in 2018. Significantly, it is shown that the proposed system serves as an integrable (3+1)-dimensional generalization of the well-studied (2+1)-dimensional dispersionless Davey-Stewartson system. This way, an interesting new example on integrability in higher dimensions is presented, with potential applications in modern mathematical physics. The work lays the foundation for future research into symmetries, conservation laws, and Hamiltonian structures, offering avenues for further exploration.
\end{abstract}





\section{Introduction}

Within the framework of partial differential equations (PDEs), integrable systems are of great interest in modern mathematics and physics \cite{ablowitz1981solitons,calogero1991certain,mason1996integrability,doktorov2007dressing}. In particular, taking into account that the vast majority of physical theories commonly accept that there are three spatial dimensions and one temporal dimension, the search for integrable systems in (3+1) dimensions becomes particularly important and challenging, cf.\ e.g.\  \cite{witten1992searching, mason1996integrability, lyx, sergyeyev2018new}. 

The paper \cite{sergyeyev2018new} gives a systematic approach for constructing (3+1)-dimensional integrable dispersionless systems using contact Lax pairs. These are nonisospectral Lax pairs associated with contact vector fields and so, at least in principle, they are amenable to the inverse scattering transform and other techniques like the dressing method \cite{doktorov2007dressing,kodama1989method,konopelchenko2002dispersionless}. 
Remarkably, the systems obtained using the method of \cite{sergyeyev2018new} admit a straightforward reduction to (2+1)-dimensional systems, which suggests that the said method allows for finding integrable (3+1)-dimensional generalizations of certain (2+1)-dimensional systems.

In this paper we implement this strategy for finding a (3+1)-dimensional integrable generalization for the dispersionless Davey-Stewartson (dDS) system, which is currently a subject of intense research (see e.g.\  \cite{yi2020dispersionless,yi2022dispersionless,go21, yi2022aml} and references therein), in view of its providing a dispersionless limit for the the dispersive Davey-Stewartson system. The latter plays a significant role in mathematical physics, as it represents a (2+1)-dimensional generalization of the nonlinear Schrödinger equation, and describes many important wave processes \cite{elena1992,ablowitz1975nonlinear,linaresDS, xzl}. 

In what follows, after a summary of relevant facts in Section \ref{SecPrelim}, we will employ the method from \cite{sergyeyev2018new} for constructing a new (3+1)-dimensional dispersionless integrable system (\ref{nuevo4D}) in Section \ref{SecNew}, and we will show that this new system is a (3+1)-dimensional generalization of dDS system (see Theorem~\ref{t-1}). We note that it could be of interest to study the symmetries, conservation laws, Hamiltonian operators, and other related structures for \eqref{nuevo4D} using e.g. the techniques from \cite{krasil2017symbolic,olver86}.\looseness=-1

\section{Preliminaries}\label{SecPrelim}
In the present paper a PDE system in four independent variables will be called a (3+1)-dimensional dispersionless system (note that dispersionless systems are also known as hydrodynamic-type systems) \cite{dunajski2015einstein,ferapontov2009hamiltonian,sergyeyev2018new,serre1994systemes} if it can be written in the form
\begin{equation}\label{dispersionless}
A_0(\mathbf u) \mathbf u_t + A_1(\mathbf u) \mathbf u_x + A_2(\mathbf u) \mathbf u_y + A_3(\mathbf u) \mathbf u_z = 0,
\end{equation}
where \(x, y, z, t\) are independent variables, \(\mathbf u = (u_1, \ldots, u_N)^T\) is an \(N\)-dimensional vector of unknown functions, and \(A_i (\mathbf u)\) are \(M \times N\) matrices whose entries depend on the components of \(\mathbf u\). Both \(M\) and \(N\) are positive integers satisfying \(M \geq N\). All functions under consideration are tacitly assumed smooth enough to ensure the validity of all computations. In what follows a (3+1)-dimensional dispersionless system will be said  to be {\em integrable} if it admits a Lax pair.\looseness=-1 

In the recent paper \cite{sergyeyev2018new} is presented a systematic approach for constructing integrable (3+1)-dimensional dispersionless systems, that we briefly review here for the readers' convenience.

Consider, following \cite{sergyeyev2018new}, the class of Lax pairs $L\chi=0, M\chi=0$, where 
\begin{equation}\label{contactlaxpair}
\begin{aligned}
    	L&=\partial_y -X_f ,\\
		M&=\partial_t -X_g ,
\end{aligned}
\end{equation}
for suitable functions \(f = f(p, \mathbf u)\) and \(g = g(p, \mathbf u)\), and with 
$$
X_h = h_p \partial_x + (p h_z - h_x )\partial_p + (h - p h_p )\partial_z.
$$
Note that \(p\) plays the role of the spectral parameter, since \(\mathbf u_p \equiv 0\) by assumption. 

These nonisospectral Lax pairs have a profound connection to contact geometry. The vector field $X_h$ 
formally 
looks exactly like 
a contact vector field corresponding to a contact Hamiltonian \(h\) on a contact 3-manifold with local coordinates \(x, z, p\) and the associated contact one-form \(dz + p dx\) (see \cite{blair2010riemannian,bravetti2017contact} and the references therein).
\vspace{0.2cm}

We state without proof the following proposition, which corresponds to Proposition 1 in \cite{sergyeyev2018new}:
\begin{proposition} \label{propZero}
A nonisospectral Lax pair of the form \eqref{contactlaxpair} is compatible if and only if the functions $f,g$ satisfy
\begin{equation}\label{zerocurvature}
f_t-g_y+\{f,g\}=0,
\end{equation}	
with the bracket $\{,\}$ defined as
$$
\{f, g\}=f_p g_x-g_p f_x-p\left(f_p g_z-g_p f_z\right)+f g_z-g f_z.
$$
\end{proposition}

Observe that equation \eqref{zerocurvature} can be rewritten as 
\begin{equation}\label{extzc}
		\sum_{i=1}^N f_{u_i}\left(u_i\right)_t-g_{u_i}\left(u_i\right)_y+\left(f_p g_{u_i}-g_p f_{u_i}\right)\left(u_i\right)_x +\left(\left(f-p f_p\right) g_{u_i}-\left(g-p g_p\right) f_{u_i}\right)\left(u_i\right)_z=0,
\end{equation}
and if the dependence of $f$ and $g$ on $p$ is rational, we can bring equation \eqref{extzc} to a common denominator. Then, vanishing of the expression in question is equivalent to that of its numerator, and the latter is achieved by equating to zero the coefficients at all powers of $p$. Consequently, Proposition \ref{propZero} allows us to produce integrable systems of 
the form \eqref{dispersionless} by 
making appropriate choices of functions $f,g$. 

Importantly, not every pair of functions $f,g$ is \textit{admissible}, in the sense that equation \eqref{zerocurvature} leads to an expression that can be transformed into a system of Cauchy–Kowalevski type by a suitable change of variables. In this sense, in \cite{sergyeyev2018new, serg18rational} it is shown that for any natural numbers \(m\) and \(n\), the pairs of Lax functions 
\begin{equation}\label{adpolfunct}
f = p^{n+1} + \sum_{j=0}^{n} v_j p^j, \quad g = p^{m+1} + \frac{m}{n} v_n p^m + \sum_{k=0}^{m-1} w_k p^k,    
\end{equation}
and 
\begin{equation}\label{adratfunct}
f = \sum_{j=1}^{n} \frac{a_i}{v_i - p}, \quad g = \sum_{k=1}^{m} \frac{b_k}{w_k - p},    
\end{equation}
are admissible, with $\mathbf u$ being in these cases  
$$
\mathbf u = (v_0, \ldots, v_n, w_0, \ldots, w_{m-1})^T
$$
and 
$$
\mathbf u = (a_1, \ldots, a_m, b_1, \ldots, b_n, v_1, \ldots, v_m, w_1, \ldots, w_n)^T,
$$
respectively.

%

Furthermore, in \cite{serg19algebraic} it is presented an example of algebraic, rather than rational, admissible pair of Lax functions not belonging to any of the classes above.

\begin{remark}\label{reduct}
As highlighted in \cite{sergyeyev2018new,serg18rational,serg19algebraic}, under the assumption \(\mathbf u_z = 0\), system \eqref{dispersionless} reduces to a (2+1)-dimensional dispersionless system. 
If \eqref{dispersionless} admits a contact Lax pair \eqref{contactlaxpair} then the system obtained from \eqref{dispersionless} by putting \(\mathbf u_z = 0\) admits a Lax pair 
involving Hamiltonian vector fields (as opposed to contact vector fields) with the Lax operators given by 
\begin{equation}\label{hamiltonianlaxpair}
		L=\partial_y -\tilde{X}_f ,\quad 
		M=\partial_t -\tilde{X}_g ,
\end{equation}
where now
$$
\tilde{X}_h = h_p \partial_x - h_x \partial_p.
$$
Detailed discussions on this type of (2+1)-dimensional integrable systems can be found e.g.\ in 
\cite{ferapontov2009hamiltonian,zakharov1994dispersionless}.	
\end{remark}

An important example of 
a (2+1)-dimensional integrable dispersionless system is given by the dispersionless Davey-Stewartson system, which is a physically relevant generalization of the dispersionless nonlinear Schrödinger equation, cf.\ e.g.\ \cite{yi2020dispersionless}. 
This system 
has 
the form:
\begin{equation}\label{dDSsystem}
	\begin{array}{r@{}l}
		U_t+2\left(U S_x\right)_x+2\left(U S_{y}\right)_{y}&=0, \\[2mm]
		S_t+S_x^2+S_{y}^2-\delta \phi&=0, \\[2mm]
		\phi_{x y}-\frac{1}{2}\left(U_{x x}+U_{y y}\right)&=0 .
	\end{array}
\end{equation}
Observe that we can eliminate $\phi$ from this system, expressing it using the second equation and substituting the result into the third one, which yields
\begin{equation}\label{dDS}
	\begin{array}{r@{}l}
		U_t+2\left(U S_x\right)_x+2\left(U S_{y}\right)_{y}&=0, \\[2mm]
		\displaystyle\frac1{\delta}(S_t+S_x^2+S_{y}^2)_{x y}-\frac{1}{2}\left(U_{x x}+U_{y y}\right)&=0 .
	\end{array}
\end{equation}

\section{Integrable (3+1)-dimensional generalization of the (2+1)-dimensional dDS system}\label{SecNew}
Consider the following system
\begin{equation}\label{nuevo4D}
	\begin{array}{rl}
		u_t = &2rv_x+vr_x+su_z-us_z+wu_x-2vw_z+s_y, \\[0.2cm]		
		v_t = &2qu_z-uq_z+sv_z-2vs_z+vw_x+wv_x+q_y,\\[0.2cm]
		
		w_y = &-2ru_x+rv_z+2vr_z+uw_z,\\[0.2cm]
		
		r_y = &ru_z+ur_z, \\[0.2cm]
		
		q_x = & 2cvu_x+cvv_z+\dfrac{q}{v}v_x, \\[0.2cm]
		
		s_x = & \dfrac{q}{v}u_x-3cvu_z-2cv_y+\dfrac{ 2cuv-2q }{v}v_z+2q_z, \\
	\end{array}
\end{equation}
where $u,v,w,r,q,s$ are unknown functions of the independent variables $x,y,z,t$, and $c\in \mathbb R$ is a constant.

%
%
%

\vspace{0.2cm}

This system can be transformed into evolutionary form, so that the Cauchy--Kowalevski theorem could be applied to it 
(so in particular the system in question is neither underdetermined nor overdetermined). 
Indeed, the coefficients of the variables $u_z, v_z, w_z, q_z, r_z, s_z$ in the previous equations constitute a nondegenerate matrix
$$
\begin{bmatrix}
	s & 0 & -2v & 0 & 0 & -u \\
	2q & s & 0 & -u & 0 & -2v \\
	0  & r & u & 0  & 2v & 0 \\
	r & 0 & 0 & 0   & u & 0 \\
	0 & cv & 0 & 0 & 0 & 0 \\
	-3cv& \dfrac{ 2cuv-2q }{v} & 0 & 2 & 0 & 0 \\
\end{bmatrix}
$$
whose determinant is 
\begin{equation}
cv\left(3cu^4v-4qu^3-16rv^3+4su^2v\right)\neq 0,
\end{equation}
so the system can be put in an evolutionary form 
with $z$ playing the role of the evolution parameter. 

\vspace{0.2cm}

Our main result establishes that system \eqref{nuevo4D} is integrable and that it is a (3+1)-dimensional generalization of \eqref{dDS}:

\begin{theorem}\label{t-1}
The (3+1)-dimensional dispersionless system \eqref{nuevo4D} admits the Lax pair
\begin{equation}\label{laxpair4D}
	\begin{aligned}
		L=\partial_y&+\dfrac{v}{p^2} \partial_x-\left(u+\dfrac{2 v}{p} \right)\partial_z-\left( u_z p-u_x+v_z-\dfrac{v_x}{p}\right) \partial_p, \\[4mm]
		M=\partial_t&-\left(2 r p+w-\dfrac{q }{p^2}-\dfrac{2 c v^2}{p^3} \right)\partial_x+\left(r p^2-s-\dfrac{2 q }{p}-\dfrac{3 c v^2}{p^2} \right)\partial_z\\
		  &-\left( r_zp^3+(w_z-r_x )p^2+( s_z-w_x) p+q_z-s_x+\dfrac{2 c  v v_z-q_x}{p}-\dfrac{2 c v v_x}{p^2} \right)\partial_p.
	\end{aligned}
\end{equation}

Moreover, for $c=-1$ the system \eqref{nuevo4D} is an integrable (3+1)-dimensional generalization of the (2+1)-dimensional dispersionless Davey-Stewartson system \eqref{dDS}.
\end{theorem}

\begin{remark}
Observe that \eqref{nuevo4D} also is a generalization of the (3+1)-dimensionl integrable system appearing in Example 1 in \cite{sergyeyev2018new}: 
if we assume $s=q=0$ and $c=0$ in \eqref{nuevo4D} we obtain
\begin{equation}\label{example1}
	\begin{array}{rl}
		u_t = &2rv_x+vr_x+wu_x-2vw_z, \\
		v_t = &vw_x+wv_x,\\
		w_y = &-2ru_x+rv_z+2vr_z+uw_z,\\
		r_y = &ru_z+ur_z, \\
	\end{array}
\end{equation}
which coincides with 
system (38) in \cite{sergyeyev2018new}.

\end{remark}

\section{Proof of Theorem \ref{t-1}}\label{SecProof}

First, observe that the Lax pair \eqref{laxpair4D} is of the form \eqref{contactlaxpair}, with the functions
\begin{equation}\label{funct_fg}
	\begin{array}{l}
		f=u+\dfrac{v}{p}, \\
		g=r p^2+w p+s+ \dfrac{q}{p}+ c \dfrac{v^2}{p^2}. 
	\end{array}
\end{equation}

Now, according to Proposition \ref{propZero}, this Lax pair is compatible if and only if 
\begin{equation}\label{zerocurvature2}
    f_t-g_y+\{f,g\}=0. 
\end{equation}
Therefore, to show that system \eqref{nuevo4D} implies the compatibility of Lax pair \eqref{laxpair4D}, it is enough to substitute \eqref{funct_fg} into \eqref{zerocurvature2}, and check that the resulting expression is equivalent to \eqref{nuevo4D} upon equating to zero the coefficients at all powers of $p$, as can be verified by straightforward algebraic manipulations.

To prove the second part of the theorem observe first that, according to Remark \ref{reduct}, if we assume
\begin{equation}\label{assumptions1}
u_z=v_z=w_z=q_z=r_z=s_z=0
\end{equation}
in system \eqref{nuevo4D} we obtain a (2+1)-dimensional integrable system. 

If we further assume, in addition to \eqref{assumptions1}, that $r=1$, system \eqref{nuevo4D} gives rise to the system
\begin{equation}\label{nuevo3D}
\begin{array}{rl}
u_t =& wu_x+s_y+2v_x,\\[0.2cm]
v_t =& vw_x+wv_x+q_y,\\[0.2cm]
w_y =& -2u_x,\\[0.2cm]
q_x =& 2cv u_x+\dfrac{q}{v}v_x,\\[0.2cm]
s_x =& -2c v_y+\dfrac{q}{v}u_x.\\[0.2cm]
\end{array}
\end{equation}

Now note that for $c=-1$ system \eqref{nuevo3D} can be reduced, by means of the following substitutions
$$
u=S_y,\quad v= \frac{1}{4}\delta U,\quad w=-2S_x,\quad q= -\frac{1}{2}\delta U S_y,\quad s=\frac{1}{2}\delta W - S_y^2,
$$
where $\delta$ is a constant and $S,U,W$ are functions of $x,y,t$,
to the system
\begin{subequations}\label{SUW}
	\begin{align}
		U_t &= -2(U S_{x})_x-2(U S_{y})_y, \label{eq2:ut}\\
		W_x &= U_y, \label{eq2:wx}\\
		S_{ty} &= -2S_x S_{xy}+\frac{1}{2}\delta W_y-2S_y S_{yy}+\frac{1}{2}\delta U_x. \label{eq2v2:s2t}
	\end{align}
\end{subequations}

We can rearrange equation \eqref{eq2v2:s2t} obtaining
\begin{equation}\label{nuev2c}
(S_{t})_y = -(S_x^2)_y-(S_y^2)_y+\frac{\delta}{2} (W_y+U_x),
\end{equation}
and differentiating \eqref{nuev2c} with respect to $x$ gives us
\begin{equation}\label{nuev3c}
    (S_{t})_{xy} = -(S_x^2)_{xy}-(S_y^2)_{xy}+\frac{\delta}{2} (W_{xy}+U_{xx}).
\end{equation}
Finally, we observe that if $S,U,$ and $W$ satisfy \eqref{SUW} then $S$ and $U$ also satisfy
the system
\begin{subequations}
	\begin{align}
		U_t &= -2(U S_{x})_x-2(U S_{y})_y, \label{nuev4a}\\
		(S_{t})_{xy} &= -(S_x^2)_{xy}-(S_y^2)_{xy}+\frac{\delta}{2} (U_{yy}+U_{xx}), \label{nuev4b}
	\end{align}
\end{subequations}
which is nothing but (\ref{dDS}). 

So, we see that system (\ref{SUW}) is an integrable generalization of (\ref{dDS}), system \eqref{nuevo3D} is an integrable generalization of \eqref{SUW}, and finally system (\ref{nuevo4D}) is an integrable (3+1)-dimensional generalization of \eqref{nuevo3D} and hence \emph{a fortiori} of \eqref{dDS}, and the result follows.

\begin{remark}
It is worth noting that the pair of functions \eqref{funct_fg} doesn't belong to any of the previously known classes of admissible pairs of rational functions \eqref{adpolfunct} and \eqref{adratfunct} yielding (3+1)-dimensional dispersionless integrable systems.
\end{remark}

\section*{Acknowledgments}
The author thanks the financial support from the {\it University of Cádiz} through the internal funding program `Plan Propio de Estímulo y Apoyo a la Investigación y Transferencia 2022/2023', and from {\it Junta de Andaluc\'ia} through the research group FQM--377.

The author would also  like to extend his sincere gratitude to A. Sergyeyev for stimulating discussions and encouragement.

\bibliographystyle{ieeetr}
\bibliography{references.bib}

\end{document}